\documentclass[aps,preprint,superscriptaddress,groupedaddress]{revtex4}
\usepackage{graphicx}
\usepackage{epsfig}
\usepackage{amssymb}
\usepackage{mathrsfs}
\usepackage{amsmath}
\usepackage{color}
\usepackage{latexsym}
\usepackage{amsfonts}
\usepackage{hyperref}
\usepackage{subfigure}
\usepackage{wasysym}
\usepackage{dcolumn}% Align table columns on decimal point
\usepackage{bm}% bold math

\begin{document}

%\definecolor{corr_16_june}{rgb}{0,0,0} % orange

\selectfont

\title{Nanoscale Dynamics of Phase Flipping in Water near its
  Hypothesized Liquid-Liquid Critical Point}

\author{T. A. \surname{Kesselring}}
\email{tobiaskesselring@ethz.ch}
\affiliation{Computational Physics, IfB, ETH Zurich, Schafmattstrasse 6,
  8093 Zurich, Switzerland} 
\author{G. \surname{Franzese}}
\affiliation{Departament de F\`{\i}sica Fonomental, Universitat de
  Barcelona, Diagonal 645, 08028 Barcelona, Spain} 
\author{S. V. \surname{Buldyrev}}
\affiliation{Department of Physics, Yeshiva University, 500 West 185th
  Street, New York, NY 10033}  
  \author{H. J. \surname{Herrmann}} 
\affiliation{Computational Physics, IfB, ETH Zurich, Schafmattstrasse 6,
  8093 Zurich, Switzerland}
\affiliation{Departamento de F\`{\i}sica, Universidade Federal do Cear\'a,
  Campus do Pici, 60451-970 Fortaleza, Cear\'a, Brazil}   
\author{H. E. \surname{Stanley}}
\affiliation{Center for Polymer Studies and Department of Physics,
  Boston University, Boston, MA 02215} 

\date{10 May 2012 --- v2-kfbhs10may.tex}

\definecolor{corr14okt}{rgb}{0,0,1} % blue
\definecolor{moved}{rgb}{0,0,0}

\maketitle

\noindent{\bf One hypothesized explanation for water's anomalies
  imagines the existence of a liquid-liquid (LL) phase transition line
  separating two liquid phases and terminating at a LL critical point.
  We simulate the classic ST2 model of water for times up to 1000 ns
  and system size up to $N=729$.  We find that for state points
  near the LL transition line, the entire system flips rapidly between
  liquid states of high and low density. Our finite-size scaling
  analysis accurately locates both the LL transition line and its
  associated LL critical point.  We test the stability of the two
  liquids with respect to the crystal and find that of the 350 systems
  simulated, only 3 of them crystallize and these 3 for the relatively
  small system size N=343 while for all other simulations the
  incipient crystallites vanish on a time scales smaller than $\approx
  100$ns.}
  
We perform extensive molecular dynamics (MD) simulations of ST2-water in
the constant-temperature, constant-pressure ensemble. We equilibrate the
system for $\approx 1000\rm{ns}$ for 127 state points in the supercooled
liquid region of water.  Pressure $P$ ranges from 190~MPa to 240~MPa,
while temperature $T$ is as low as $T=$230~K at high $P$, and 244~K at
low $P$. We make 624 different simulations, 341 as long as 1000 ns, and
for four system sizes, $N$ = 216 (80 state points), 343 (75 state
points), 512 (44 state points), and 729 molecules (46 state points).
For the majority of state points studied we average our results over
several ($\leq 11$) independent runs.  We interpolate our data along
isobars using the histogram reweighting method
\cite{Panagiotopoulos2000}.  For $P\gtrsim 200$~MPa, we find that the
density $\rho$ decreases sharply within a narrow temperature range,
while at lower $P$ it falls off with $T$ continuously. This behavior is
consistent with a discontinuous phase transition at high-$P$ between a
high-density liquid (HDL) and a low-density liquid (LDL) ending in a
liquid-liquid (LL) critical point at lower $P$
(Fig.~\ref{overview_runs}a).

This LL critical point was hypothesized \cite{llcp} based on
studies of the ST2 model, and subsequently studied in detail by many
others using, in addition to ST2 \cite{Poole2005,liu2009},  TIP5P
\cite{4A}, TIP4P \cite{Dario},  TIP4P-Ew \cite{Dietmar-Paschek:2008vl} and TIP4P/2005
\cite{abascal2010} as well as coarse-grained models
\cite{FMS2003,franzese,starr}.  The existence of the LL 
critical point allows one to understand X-ray spectroscopy results
\cite{Tokushima2008,Nilsson2011,Wikfeldt2011}, and explains the
increasing correlation length in bulk water upon cooling as found
experimentally \cite{Huang2010} and the hysteresis effects
\cite{Zhang:2011uq}.  Holten et al. \cite{Holten2012,15B} reviews
available experimental information and shows that the assumption
of a LL critical point in supercooled water provides an accurate account
on the experimental thermodynamic properties.

Abrupt changes in the global density $\rho$ are related to the
appearance of different local structures. Among various parameters
describing the local structures we identify $d_3$
\cite{Ghiringhelli2008} and $\psi_3$, defined in the Methods Section, as good
quantities to distinguish the LDL and the HDL phase and the best quantities
to distinguish them from ice.  The average values of $\psi_3$ of the two
phases differ by about 50\%, the LDL phase being characterized by
greater order in the second shell than in the HDL phase.

Liu et al. \cite{liu2009}, using histogram reweighed Monte Carlo methods in the
 grand canonical ensemble for only one but quite large system size, find
 an order parameter distribution function consistent with a critical
 point belonging to the universality class of a 3 dimensional Ising
 model. In Ref. \cite{Limmer:2011uq} Limmer and Chandler question this result using the
 umbrella sampling method to evaluate the free energy landscape of the
 ST2 model near a single state point and for a single system size
 ($N=216$). They find two minima in the free energy landscape: one for
 liquids and one for crystalline structure. They do not find a third
 minimum corresponding to the LDL and conclude that the LDL does not
 exist as a metastable state, but only as a transitional state from HDL
 to crystal. However, Sciortino et al. in \cite{Sciortino2011} show, with an
 implementation for $200\geq N \geq 327$ of the umbrella sampling that
 guarantees very high resolution in the exploration of the free energy
 landscape, the presence of the minimum corresponding to the LDL state
 metastable with respect to the crystal, reconfirming the results of Ref.
\cite{liu2009} and at variance with Ref. \cite{Limmer:2011uq}. To contribute to the discussion, we
 present here a finite size scaling analysis of results from extremely
 long ($1\mu$s) MD simulations. We find that 1) LDL is a genuine liquid
 state, metastable with respect to the crystal, 2) LDL and HDL are
 separated by a first-order phase transition line ending in a critical
 point, 3) LDL has relaxation times that exceed $1\mu$s at
 temperatures below the
 LDL-HDL coexistence at low pressure, 4) the results are robust with
 respect to the finite size scaling analysis and show that the LDL-HDL
 critical point belongs to the 3 dimensional Ising model.

\medskip
\bigskip
%\medskip
\noindent{\bf RESULTS}

To show that the LL phase transition exists in the thermodynamic limit,
we perform a finite-size analysis along isobars within the supercooled
liquid region.  For this purpose, we calculate the Challa-Landau-Binder
parameter $\Pi\equiv 1-\langle\rho^4\rangle/3\langle\rho^2\rangle^2$ for
the bimodality of the density distribution function, ${\cal D}(\rho)$
\cite{Challa1986,Franzese2000}.  When ${\cal D}(\rho)$ is unimodal,
$\Pi$ adopts the value $2/3$ in the thermodynamic limit $N\rightarrow
\infty$, while $\Pi<2/3$ when ${\cal D}(\rho)$ is bimodal, since two
phases coexist (Fig.~\ref{fig:binder}b).

However, for a finite system $\Pi<2/3$ whenever ${\cal D}(\rho)$
deviates from a delta function. This occurs in the region of the phase
diagram where, for a finite system, the isothermal
compressibility, $K_T$, has a maximum, i.e., along a locus in the
$P$--$T$ plane that includes (i) the discontinuous (in the
thermodynamic limit) phase transition at
$P>P_c$, the LL critical pressure, (ii) the effective LL critical point
at $P_c(N)$, where the discontinuity vanishes, and (iii) a line for
$P<P_c$ that emanates from the LL critical point into the supercritical
region.  Near $P_c$ this line follows the locus of maxima of the
correlation length, known as the Widom line \cite{Xu2005}, and deviates
from it at lower $P$ \cite{FS2007}.

The finite-size behavior of $\Pi$ allows us to distinguish whether an
isobar is above or below $P_c$ \cite{Challa1986,Franzese2000}
(Fig.~\ref{fig:binder}b).  When isobars cross the Widom line ($P<P_c$),
$\Pi$ displays a minimum $\Pi_{\rm min}$ (inset in Fig.1b) that in leading order
approaches 2/3 linearly with $1/N$. When ${\cal D}(\rho)$ consists of
two Gaussians of equal weight, i.e. at the coexistence line for $P
\gtrsim P_c$, $\Pi_{\rm min}$ approaches, also linearly with $1/N$,
another limiting value $\Pi\rightarrow 2/3-(\rho_{\rm H}^2-\rho_{\rm
  L}^2)^2/ [3(\rho_{\rm H}^2+\rho_{\rm L}^2)^2]$ where $\rho_{\rm
  H}=\rho_{\rm H}(P)$ and $\rho_{\rm L}=\rho_{\rm L}(P)$ are the
densities of the two coexisting phases \cite{Challa1986}. This limiting
value progressively decreases as $P$ increase above $P_c$, since
$\rho_{\rm H}-\rho_{\rm L}$ increases at coexistence as $(P-P_c)^\beta$,
where $\beta\approx 0.3$ is the critical exponent of the 3d Ising
universality class \cite{Holten2012}.

To ensure that the system is in thermal equilibrium, we calculate the
correlation time for the first maximum $k_1$ of the oxygen-oxygen
intermediate scattering function $S_{\rm OO}(k,t)$, as defined in the
Methods Section.  While correlation times in the HDL phase are very
short ($\approx 0.01$~ns), they become of the order of 100~ns in the LDL
phase, implying that simulations of less than 1~$\mu$s are likely
affected by poor statistical sampling (Fig.~\ref{fig:sq}). For
temperatures above the line $T_g$ in Fig.~\ref{overview_runs}a,
correlation times are smaller than 100~ns and we can equilibrate the
system within our simulation time.

Figure \ref{close-to-coex}a shows a typical example of a simulation near
the critical point for $N=343$ molecules at $P=215$~MPa and
$T=244$~K. Here the system exhibits phase flipping between LDL and HDL,
with the life-time of each phase distributed from $\approx 20$ ns to
$\approx 300$ns.  This nanoscale phase flipping results in a bimodal
density distribution (Fig.~\ref{close-to-coex}b) and is observed for all
temperatures and pressures around the LL phase transition in a region
that shrinks with growing system size.

\medskip
\bigskip
\noindent{\bf DISCUSSION}

To estimate the critical exponents of the LL critical point we next
investigate the distribution of the order parameter $M$ of the LL phase
transition.  As for the liquid-gas phase transition \cite{Wilding1997},
the order parameter is not simply the density, but a linear combination
of the density with another observable \cite{Bertrand2011}. Here we
choose the linear combination of density and energy $M \equiv \rho+sE$
\cite{Wilding1997} and find that it follows, as expected, the behavior
of a liquid in the universality class of the three dimensional (3d)
Ising model, as is also the case for the liquid-gas transition
(Fig. \ref{fig:orderparam_distr_fkt}). At $P=205$~MPa~ the difference
between the maxima and the central minimum of the order parameter
distribution is smaller than for the 3d Ising case. At $P=210$~MPa~ it
is larger and the critical point therefore seems to be in between,
consistent with the conclusion obtained from the analysis of $\Pi$.  We
get the best fit of the order parameter distribution function at a
pressure of $P=206\pm3$~MPa~ and a temperature of $T=246 \pm 1$~K.

The same analysis for $N=512$ and 729 yields estimates, consistent with
$N=343$, of the LL critical point to be $P_c=208\pm3$~MPa and
$T_c=246\pm1$~K (Fig.~\ref{fig:orderparam_distr_fkt}b, c).  The finite
size scaling of the amplitudes of the order parameter distribution $A
\sim L^{\beta/ \nu}$ is consistent with the behavior predicted for the
3d Ising universality class with $\beta / \nu \approx 0.518$
\cite{Wilding1997} and strong corrections to scaling for $N \lesssim
343$ (Fig. \ref{fig:orderparam_distr_fkt}d).

Finally, we investigate also the possibility of spontaneous crystal
nucleation in the LDL phase using the structural order parameter $d_3$
\cite{Ghiringhelli2008}. At temperatures below the region of phase
flipping, the samples sometimes form large crystallites filling up to
10\% of the system volume. Their structure exhibits a mixture of cubic
and hexagonal symmetry. However, in approximately 99\% of simulations
these unstable crystallites vanish within the simulation time of
1~$\mu$s, showing that the free-energy barrier for the crystallization
process is significantly larger than $k_BT$ in the LDL phase
(Fig.~\ref{fig:cryst}).  We observe irreversible crystallization in only
3 out of 350 (1~$\mu$s)--runs, for only $N=343$ and all corresponding to
state points near the LL critical point
(Fig.~\ref{overview_runs}a). This is consistent with the general result
that a metastable fluid-fluid phase transition favors the
crystallization process in its vicinity \cite{tenWolde1997}.  We did not
observe any crystallization events for $N=512$ and $N=729$ although the
total simulation time for these systems is comparable to that of
$N=343$. The fact that the crystallization rate is not increasing with
system size is evidence that LDL is the genuine metastable phase with
respect to the stable crystal phase.

In conclusion, we use new methods to investigate both the statics and
dynamics of deeply supercooled ST2-water.  Specifically, we analyze
static quantities (density and potential energy)
 using the framework of finite-size scaling theory,
and we analyze the dynamic structure factor
over three orders of magnitude of
time scales, from 1 to 1000~ns.  We find definitive evidence of a first
order LL phase transition line between two genuine phases that are each
metastable with respect to a liquid. The phase transition line
terminates in a LL critical point, and the exponents associated with
this LL critical point are indistinguishable from those expected for a three-dimensional
lattice-gas model which is used to describe the liquid-vapor critical
point.

\medskip
\bigskip
\noindent{\bf METHODS}

We performed MD simulations in the $NPT$ ensemble using the Stillinger
and Rahman \cite{ST2} five-point water model ST2, consisting of five
particles interacting through electrostatic and Lennard-Jones forces
with a cutoff of 7.8~\AA.  The pressure was not adjusted to correct for
the effects of the Lennard-Jones cutoff, since it would originate from
mean field calculations, which become rather poor near a critical point.

We apply the {\it Shake\/} algorithm to constrain the particles inside
each molecule. The constant pressure is imposed by a Berendsen barostat,
and a Nos\'e-Hoover thermostat is applied to ensure constant temperature
\cite{Allen1987}. Periodic boundary conditions have been implemented.

For the simulations we used the following protocol consisting of three
steps: (1) For any given density, a constant volume simulation is
performed at $T=300$~K during 1~ns (first pre-run). (2) The ensemble is
then changed to $NPT$ by adding the Berendsen barostat with the desired
pressure and the temperature is reduced to $T=265$~K, ensuring that the
system reaches the HDL phase after $1$~ns of equilibration (second
pre-run).  (3) After these two pre-runs the system is quenched to the
desired temperature, from which the first $100-200$~ns are removed as
thermalization time. The choice of the thermalization time will be
discussed next.

To decide whether the equilibration time is sufficient, we perform two
steps. First, we inspect the time series of energy and density to
discard the possibility of spontaneous crystallization. In all our $NPT$
simulations we observed only three crystallization events ($\approx 1\%$
of total number of runs) all of them in systems with the smallest size
($N=343$ molecules).  We use them as a reference for the crystal. In a
second step we measure the correlation time using the intermediate
scattering function.

 MD simulations are performed for a finite numbers of state points
 (Fig. \ref{fig:binder}a). We use then histogram reweighting method \cite{Panagiotopoulos2000} to complement the
 statistics of each state point with the information from nearby
 state points. Histogram reweighting \cite{Ferrenberg} is a method that combines
 the overlapping histograms of quantities calculated at close-enough
 state points, reweighting them with an appropriate factor that takes
 into account the difference in thermodynamic parameters. It is a
 powerful method that allows to calculate the observables for a
 continuous range of thermodynamic parameters within those directly
 simulated.

The order parameter $M\equiv \rho+sE$ is obtained from the distribution
in the density--energy plane (Fig. \ref{fig:orderparam_distr_fkt}e), by
integrating it with a delta-function $\delta(M-\rho-sE)$. We select the
value of $s$ for which the distribution of $M$ best fits the
distribution of the order parameter for the 3d Ising universality class.
The main effect found when changing $s$ is a small shift in the
estimated critical temperature $T_C$ of about $0.1$~K, which is less
than the error of $0.5$~K originating from the histogram reweighting.

The oxygen-oxygen intermediate scattering function $S_{\rm OO}({\bf
  k},t)$ can be used to distinguish between phases of different
structure, such as LDL and HDL. We also use it to estimate the
correlation time.  It is defined as
\begin{eqnarray}
S_{\rm OO}( {\bf k}, t) \equiv \frac{1}{N}\left< \sum_{\ell,m}^N \exp(i
  {\bf k} \cdot \left[ {\bf r}_\ell(t^\prime)-{\bf
      r}_m(t^\prime+t)\right])\right>_{t^\prime}, 
\label{eqn:Intermediate_structure_fkt}
\end{eqnarray}
where $\left< . . . \right>_{t^\prime}$ denotes averaging over the
simulation time $t^\prime$, ${\bf r}_\ell(t^\prime)$ is the position of the
oxygen of molecule $l$ at time $t^\prime$, ${\bf k}$ is the wave vector
and $k$ is its magnitude $|{\bf k}|$.  $S_{\rm OO}({\bf k},t)$ describes
the time evolution of the spatial correlation along the wave vector
${\bf k}$. Since the system has periodic boundaries, the components of
${\bf k}$ have discrete values $2\pi j/L$, where $L$ is the length of
the simulation box and $j=1,2,...$.  We define 
$
S_{\rm OO}(k,t)\equiv \langle S({\bf k},t)\rangle_j,
$
where average is taken over all vectors ${\bf k}$ with magnitude $k$ belonging to $j$th spherical bin $\pi(j-1/2)/L\leq k <\pi(j+1/2)/L$, for $j=2,3,...300$.

The temporal decay of $S_{\rm OO}(k,t)$ is characterized by two
relaxation times: (i) a short time, $\tau_\beta$, after which $S_{\rm
  OO}(k,t)$ reaches a plateau $S_{\rm OO}(k,\tau_\beta)$ corresponding
to the bouncing of the particles inside the cages formed by their
neighbors, and (ii) a long time, $\tau_\alpha$, corresponding to a
particle escaping from its cage and diffusing away from its initial
position. We define the correlation time $\tau=\tau_\alpha$ as the time
for which $C_{\rm OO}(k,\tau)\equiv S_{\rm OO}(k,\tau)/S_{\rm
  OO}(k,\tau_\beta)=1/e$, where $C_{\rm OO}(k,\tau)$ is the structural
correlation function (Fig.~\ref{fig:sq}).

We define the bond order parameter $d_3$ following
Ref.~\cite{Ghiringhelli2008}. The quantity $d_3(i,j)$ characterizes the
bond between molecules $i$ and $j$ and is designed to distinguish
between a fluid and a diamond structure.  It uses the $Y_{3}^m$
spherical harmonics to identify the tetragonal symmetry of the diamond
structure. In general, each molecule is characterized by a vector ${\bf
  q}_{\ell}^i$ in the $(4\ell+2)$--dimensional Euclidean space with components
${\cal R}e(q_{\ell,m}^i)$ and ${\cal I}m(q_{\ell,m}^i)$ ($m=-\ell,
...,-1,0,1,....,\ell$), with
$$q_{\ell,m}^i\equiv \frac{1}{4} \sum_{j \in n_i} Y_\ell^m
(\varphi_{ij},\vartheta_{ij}), \ \ \ -\ell \leq m \leq \ell.$$
If molecule $j$ belongs to the first coordination shell $n_i$ (shell
of four nearest neighbors) of molecule $i$, we define $d_3(i,j)$ as the
cosine of the angle between two vectors ${\bf q}_3^i$ and ${\bf q}_3^j$
characterizing the first coordination shells of molecules $j$ and
$i$, respectively:
\begin{eqnarray}
d_3(i,j)\equiv \frac{({\bf q}_3^i\cdot {\bf q}_3^j)}{|{\bf q}_3^i||{\bf
    q}_3^j|}  
\label{d3}
\end{eqnarray}
where 
$$
({\bf q}_3^i\cdot {\bf q}_3^j)\equiv \sum_{m=-\ell}^\ell({\cal
  R}e~q_{\ell,m}^i{\cal R}e~q_{\ell,m}^j+{\cal I}m~q_{\ell,m}^i{\cal
  I}m~q_{\ell,m}^j), 
$$
and  $|{\bf q}_3^i|\equiv \sqrt{({\bf q}_3^i\cdot {\bf q}_3^i)}$. 

In a perfect diamond crystal $d_3(i,j)=-1$ for all bonds, while for
a graphite crystal $d_3(i,j)=-1$ only for bonds connecting atoms in the
same layer. For bonds connecting atoms in different layers
$d_3(i,j)=-1/9$. Thus in graphite each atom has three out of four bonds
having $d_3(i,j)=-1$. In our simulations, the spontaneously grown
crystals have many defects, with different parts of the crystals
following diamond or graphite patterns (Fig.~\ref{fig:d3ij}a). Therefore, we consider a
molecule in a crystal to have either three or four bonds with $d_3(i,j)<
d_c=-0.87$, where the value of $d_c=-0.87$ is selected as two standard
deviations from the peak of the crystal histogram corresponding to
$d_3=-1$.  This is exactly the same criterion to
specify molecules in the crystal state as in
Ref.~\cite{Ghiringhelli2008}. 
We find separate crystallites using the
percolation criterion, i.e., two molecules satisfying the crystalline
criterion belong to the same crystallite if they belong to the first
coordination shell of each other (Fig. \ref{fig:cryst}).

We finally observe that 
by defining $\psi_3(i)\equiv {1\over 4}\sum_{j=1}^4d_3(i,j)$ as the
average of $d_3$ over the four bonds of each molecule, we introduce a
single-molecule structural parameter that also can be used to
distinguish among the HDL, the LDL and the crystal phase (Fig.~\ref{fig:d3ij}b).

\clearpage

%\bibliographystyle{naturemag}
%\bibliography{water,water2}

\bigskip

\medskip
\bigskip
\noindent{\bf ACKNOWLEDGEMENTS}

\noindent  We thank D. T. Limmer and D. Chandler, Y. Liu,
A. Z. Panagiotopoulos, P. Debenedetti, F. Sciortino, I.  Saika-Voivod
and P. H. Poole for sharing their results, obtained using approaches
different from ours but also addressing the question of the hypothesized
existence of a LL phase transition line and an associated LL critical
point.~\cite{Limmer:2011uq,Poole2011a,Sciortino2011}.
 We also thank S.-H.  Chen, P. H. Poole, and F. Sciortino for a
critical reading of the manuscript and for helpful suggestions.  GF
thanks Ministerio de Ciencia e Innovaci\'on-Fondo Europeo de Desarrollo
Regional (Spain) Grant FIS2009-10210 for support.  SVB acknowledges the
partial support of this research through the Dr.  Bernard W. Gamson
Computational Science Center at Yeshiva College and through the
Departament d'Universitats, Recerca i Societat de la Informaci\'o de la
Generalitat de Catalunya.  HES thanks the NSF Chemistry Division for
support (grants CHE 0911389 and CHE 0908218).
\newpage

\medskip
\bigskip
\noindent{\bf AUTHOR CONTRIBUTIONS}

T.K performed the simulations. T.K., S.B. and G.F evaluated the data. T.K., S.B., G.F., H.H. and E.S. wrote the paper. H.H and E.S supervised the project.

\medskip
\bigskip
\noindent{\bf ADDITIONAL INFORMATION}

The authors declare no competing financial interests.

\medskip
\bigskip
\noindent{\bf FIGURE LEGENDS}

\begin{figure}[h t b]
\caption{
\label{overview_runs} 
\label{fig:binder} 
{\bf Phase diagram and
    finite size scaling analysis to locate the line of liquid-liquid
    (LL) phase transitions.}  
  (a) State points in the $P$--$T$ diagram
  simulated. Different symbols correspond to different sizes $N$.  The
  high-$T$ (red) region exhibits HDL-like states and the low-$T$ (blue)
  region LDL-like states.  In the intermediate (violet) region we
  observe flipping between HDL-like and LDL-like states. Below the black
  line correlation times are larger than $100$~ns, while above they are
  smaller. Equilibrium is attained within reasonable simulation
  times. The white region, denoted CP, is our estimate of the location
  of the LL critical point in the thermodynamic limit.  (b) Finite-size
  analysis of $\Pi_{\rm min}$ along isobars crossing the discontinuous
  LL phase transition (violet at high $P$) and the Widom line
  (within the violet region at low $P$).  At $P=190$~MPa, $\Pi_{\rm min}$ 
  approaches $2/3$ when $N\rightarrow \infty$, indicating that
  the density distribution is unimodal and that one crosses the Widom
  line, and not the line of discontinuous phase transition. At
  $P=200$~MPa, $\Pi_{\rm min}$ approaches $\approx 2/3-0.001$,
  consistent within its error bar with the value expected at coexistence
  \cite{Challa1986}.  At $P=210$~MPa, $\Pi_{\rm min}$ tends to a smaller
  value clearly excluding 2/3 and therefore the distribution ${\cal
  D}(\rho)$ is bimodal, that is the fingerprint of a discontinuous LL
  phase transition. $\Pi_{\rm min}$ depends linearly on $1/N$ to the
  leading order, displaying deviations only for the smallest size
  $N=216$.  The inset shows $\Pi$ along the isobar at $P=200$~MPa as a
  function of $T$ for all four system sizes (from bottom to top:
  $N=216$, 343, 512, 729) displaying a clear minimum $\Pi_{\rm
  min}$. Lines are interpolations obtained using histogram reweighting
  for up to eleven independent simulations of length 1~$\mu$.}
\end{figure}

\begin{figure}[h t b]
\caption{\label{fig:sq} {\bf Definition of the correlation time $\tau_0$
    using the intermediate scattering function.} The correlation time
  $\tau_0$ is calculated using the correlation function $C_{\rm
    OO}(k,t)$ of the intermediate scattering function of the oxygen
  atoms $S_{\rm OO}(k,t)$.  For the $k$ vectors corresponding to the
  first three maxima $k_1$, $k_2$ and $k_3$ (marked in red, blue and
  green in the inset), we calculate the evolution of the correlation
  function $C_{\rm OO}(k_i,\tau)$. We then define the correlation time
  as the time for which $C_{\rm OO}(k_i, \tau)$ decreases to $1/e$ for
  the slowest of the $k_i$ vectors. For nearly all the state points $k_1$ has been
  the vector for which this decrease has been the slowest.
  10--100~ns for the LDL phase, so we can equilibrate this phase in our
  simulations of about 1000~ns.  Data are for a system of $N=343$
  molecules at pressure $P=210\rm{~MPa}$ and temperatures (from left to
  right) $T=244\rm{~K}$, $243\rm{~K}$, in the LDL phase, and
  $242\rm{~K}$ below the $T_g$ line of Fig.~\ref{fig:binder}a.}
\end{figure}

\begin{figure}%[h t b]
\caption{\label{close-to-coex} {\bf Phase flipping between LDL and HDL
    at coexistence.} (a) The 1~$\mu$s time series shows how frequently,
  at constant $P=215$~MPa~ and $T=244$~K, $N=343$ ST2-water molecules
  switch from LDL-like to HDL-like states. (b) The histogram for the
  sampled density values, in arbitrary units, after discarding the first
  100~ns of the 1~$\mu$s time series.  For LDL-like states $\rho\approx
  (0.89\pm 0.01)$~g/cm$^3$ and for HDL-like states $\rho\approx (1.02\pm
  0.03)$~g/cm$^3$ corresponding to a difference of $\approx 13\%$ in density.
Dashed
  lines are Gaussian best fits of the histogram around the two maxima.}
\end{figure}

\begin{figure}[h t b]
\caption{\label{fig:orderparam_distr_fkt} {\bf The liquid-liquid
    critical point falls into the same universality class as the
    liquid-gas critical point.}  (a) The distribution function of the
  rescaled order parameter $x \equiv A(M-M_c)$ where $M \equiv \rho+sE$
  with $s=27.6 \rm{\frac{g/cm^3}{KJ/mol}}$, follows for
  $P=(206\pm3)$~MPa and $T=(246\pm1)$~K (triangles) the order parameter
  distribution function of the 3d Ising model (solid line)
  \cite{Hilfer1995}.  The data are from histogram reweighting of $N=343$
  molecules at $P=205$~MPa~ and $T=246.6$~K~ (squares), $P=206$~MPa~ and
  $T=246$~K~ (triangles) and $P=210$~MPa~ and $T=245.1$~K~ (circles).
  We repeat the analysis for (b) $N=512$ and (c) $N=729$.  (d) For large
  sizes the amplitude $A$ (triangles) scales as $A\sim L^{\beta/\nu}$,
  where $\beta/\nu\approx 0.52$, as in the 3d Ising universality class
  \cite{Wilding1997}. For $N \lesssim 343$ corrections to scaling are
  strong.  (e) Contour plot of the distribution of states in the
  density-energy plane, with red corresponding to the highest values and
  blue to the lowest. The distribution of the order parameter $M\equiv
  \rho+sE$ is obtained from this two-dimensional distribution by
  integrating it with a delta-function $\delta(M-\rho-sE)$. We select
  the value of $s$ for which the distribution of $M$ best fits the
  distribution of the order parameter for the 3d Ising universality
  class.}
\end{figure}

\begin{figure}
\caption{\label{fig:cryst} {\bf Example of a simulation where the
    largest crystallite grows up to 35 molecules and then vanishes in a
    system having $N=343$ molecules at a pressure of $P=200\rm{~MPa}$
    and $T=246\rm{~K}$.}  A molecule $i$ is considered to belong to a
  crystal if $d_3(i,j)\leq d_c=-0.87$ for three out of its four bonds
  with nearest neighbors $j$.}
\end{figure}

\begin{figure}
\caption{{\bf The equilibrium probability distributions of two different
    structural parameters that allow us to distinguish among three
    structures: HDL, LDL, and crystal.}  (a) For the smallest size
  simulated ($N=343$), in the vicinity of the LL critical point in
  $\approx 1\%$ of the runs our system spontaneously crystallizes,
  forming a structures with diamond and graphite patterns and with many
  defects. This structure is well characterized by the probability
  distribution ${\cal D}(d_3)$ (line with full dots) of the parameter
  $d_3$, displaying a large maxima close to $d_3=-1$, the value that
  corresponds to the perfect diamond crystal. For the sake of comparison
  with the other cases, we divide ${\cal D}(d_3)$ of the crystal by 10.
  In the $99\%$ of our simulations we find distributions ${\cal D}(d_3)$
  as those presented here for $P=215$~MPa, a pressure above the LL
  crital point pressure, and decreasing $T$ (from right to left). ${\cal
    D}(d_3)$ shows an abrupt change when crossing the first-order LL
  phase transition region at $T\approx 244$~K.  In particular, ${\cal
    D}(d_3)$ displays a pronounced shoulder at higher $d_3$ in the LDL
  phase, and is very different from the crystal case.  The arrow marks
  the value $d_c=-0.87$ selected as two standard deviations from the
  peak of the crystal histogram corresponding to $d_3=-1$ as in
  Ref.~\cite{Ghiringhelli2008}.  (b) The equilibrium probability
  distribution of the single-molecule parameter $\psi_3$ also
  distinguishes among HDL, LDL and crystal structures.  The average
  values for fluid phases are $\psi_3=-0.34\pm 0.19$ in the HDL phase
  and $\psi_3=-0.57\pm 0.16$ in the LDL phase.}
\label{fig:d3ij} 
\end{figure}

%\newpage
%\includegraphics[width=1\textwidth]{overv_u_binder.pdf}
%\begin{center}
%\large{\bf FIG. 1}

%\newpage

%\includegraphics[width=1\textwidth]{sq_diff_T.pdf}

%\large{\bf FIG. 2}

%\newpage

%\includegraphics[width=1\textwidth]{test_100_1000.pdf}

%\large{\bf FIG. 3}

%\newpage

%\includegraphics[width=1\textwidth]{OPDF_all.pdf}

%\large{\bf FIG. 4}

%\newpage

%\includegraphics[width=1\textwidth]{cryst.pdf}

%\large{\bf FIG. 5}

%\newpage

%\includegraphics[width=1\textwidth]{fig6.pdf}
%\large{\bf FIG. 6}

%
%\end{center}

\end{document}